\documentclass[12pt,preprint]{aastex}
\begin{document}

\def\wisk#1{\ifmmode{#1}\else{$#1$}\fi}

\def\lt     {\wisk{<}}
\def\gt     {\wisk{>}}
\def\le     {\wisk{_<\atop^=}}
\def\ge     {\wisk{_>\atop^=}}
\def\lsim   {\wisk{_<\atop^{\sim}}}
\def\gsim   {\wisk{_>\atop^{\sim}}}
\def\kms    {\wisk{{\rm ~km~s^{-1}}}}
\def\Lsun   {\wisk{{\rm L_\odot}}}
\def\Zsun   {\wisk{{\rm Z_\odot}}}
\def\Msun   {\wisk{{\rm M_\odot}}}
\def\um     {$\mu$m}
\def\mic     {\mu{\rm m}}
\def\sig    {\wisk{\sigma}}
\def\etal   {{\sl et~al.\ }}
\def\eg     {{\it e.g.\ }}
 \def\ie     {{\it i.e.\ }}
\def\bsl    {\wisk{\backslash}}
\def\by     {\wisk{\times}}
\def\half {\wisk{\frac{1}{2}}}
\def\third {\wisk{\frac{1}{3}}}
\def\nwm2sr {\wisk{\rm nW/m^2/sr\ }}
\def\nw2m4sr {\wisk{\rm nW^2/m^4/sr\ }}

\title{Demonstrating the negligible contribution of optical ACS/{\it HST} galaxies to source-subtracted cosmic infrared background 
fluctuations in deep IRAC/{\it Spitzer} images.}

\author{
A. Kashlinsky\altaffilmark{1,2,4}, R. G. Arendt\altaffilmark{1,2,4},
J. Mather \altaffilmark{1,3,4}, S. H. Moseley \altaffilmark{1,3,4} }
\altaffiltext{1}{Observational Cosmology Laboratory, Code 665,
Goddard Space Flight Center, Greenbelt MD 20771}
\altaffiltext{2}{SSAI} \altaffiltext{3}{NASA}
\altaffiltext{4}{e--mail: kashlinsky@milkyway.gsfc.nasa.gov, arendt@milkyway.gsfc.nasa.gov, mather@milkyway.gsfc.nasa.gov, moseley@milkyway.gsfc.nasa.gov}

\begin{abstract}
We study the possible contribution of optical galaxies detected with the {\it Hubble} ACS instrument to the near-IR cosmic infrared (CIB) fluctuations in deep {\it Spitzer} images. The {\it Spitzer} data used in this analysis are obtained in the course of the GOODS project from which we select four independent $10^\prime\times10^\prime$ regions observed at both 3.6 and 4.5 \um.  ACS source catalogs for all of these areas are used to construct maps containing only their emissions in the ACS $B, V, i, z$-bands. We find that deep Spitzer data exhibit CIB fluctuations remaining after removal of foreground galaxies of a very different clustering pattern at both 3.6 and 4.5 \um\  than the ACS galaxies could contribute. We also find that there are very good correlations between the ACS galaxies and the {\it removed} galaxies in the Spitzer maps, but practically no correlations remain with the residual Spitzer maps used to identify the CIB fluctuations. These contributions become negligible on larger scales used to probe the CIB fluctuations arising from clustering. This means that the ACS galaxies cannot contribute to the large-scale CIB fluctuations found in the residual Spitzer data. The absence of their contributions also means  that the CIB fluctuations arise at $z\gsim 7.5$ as the Lyman break of their sources must be redshifted past the longest ACS band, or the fluctuations have to originate in the more local but extremely low luminosity galaxies.
\end{abstract}
\keywords{cosmology: observations - diffuse radiation - early
Universe}

\section{Introduction}

The cosmic infrared background (CIB) is comprised of emission from luminous
objects from all epochs in the Universe, including by those
too faint for individual detection by current telescopic studies (see Kashlinsky 2005
for review). A particularly interesting contributor to CIB comes
from the first stars epoch (Santos et al 2002), which may have left
measurable CIB fluctuations (Kashlinsky et al 2004, Cooray et al
2004). An attempt to measure this CIB fluctuations component was
made recently by Kashlinsky et al (2005, 2007a,b - hereafter
KAMM1, KAMM2, KAMM3) using deep IRAC {\it Spitzer} images at 3.6
to 8 \um. The first results used a $\sim$ 10 hr
exposure of a $5^\prime\times10^\prime$ field from Fazio et al (2004) and, after
removing foreground galaxies to $m_{\rm AB} \sim $25 (at 3.6 $\mu$m), detected an
excess CIB fluctuations component over that from the remaining
``ordinary" galaxies at $\gsim 0.5^\prime$ (KAMM1). KAMM1 showed that the fluctuations do not arise from
zodiacal light, Galactic cirrus or instrumental noise or artifacts (see also KAMM2)
and that the residual maps do not correlate with the removed
sources. A follow-up study by KAMM2 used much
deeper GOODS data (Dickinson et al 2003) with $\gsim 20$ hr exposures in two 
independent sky areas observed at two epochs. The new data in four independent fields allowed
KAMM2 a better removal of foreground galaxies, i.e. until a fixed floor of the
residual shot noise contribution to the power spectrum, $P_{\rm
SN}$, was reached. At the same level of $P_{\rm
SN}$ all regions have the same large-scale CIB fluctuations, consistent with their cosmological origin. The new measurements allowed a firmer interpretation of the cosmological
sources generating these CIB fluctuations: they must have at most
a fairly low shot noise component and at the same time produce a
significant CIB fluctuations component at $\gsim 0.5^\prime$
arising from clustering (KAMM3). 

Recently, Cooray et al (2007) attempted to demonstrate that the
signal detected by KAMM1 at 3.6 \um\ originates from optical galaxies detected
by the {\it Hubble} ACS instrument.  In one
of the GOODS regions they blanked the pixels in
the IRAC maps corresponding to the optically identified ACS
sources and, from the remaining $\sim 20\%$ of the map, evaluated
the power spectrum of the diffuse emission claiming that it is
significantly attenuated when the ACS sources are removed. It was pointed out by Kashlinsky (2007) that
such analysis is flawed: for such deeply cut maps one cannot
evaluate clustering properties using Fourier transforms because
the basis functions are no longer even approximately orthogonal (e.g. Gorski
1994). Kashlinsky (2007) further demonstrates that when the correlation
function, $C(\theta)$, which is immune to masking, is computed
instead for the Cooray et al maps it does not depend, within the statistical uncertainties, on whether or not ACS sources are removed.

In this {\it Letter} we present the results from our study of
whether the optically detected ACS galaxies can explain the CIB
fluctuations detected by KAMM. 
Because the large--scale fluctuations do not have a high S/N for the longest IRAC channels, where the instrument noise is significant, we restricted our analysis to the 3.6 and 4.5 \um\ IRAC channels. We note that any explanation of the fluctuations must account for at least the measured
amplitude {\it and} slope at {\it both} 3.6 and 4.5 \um\ and at the right level of the shot noise (KAMM3).
We show that the galaxies detected in the ACS GOODS survey cannot contribute significantly to the CIB fluctuations detected by KAMM because they produce diffuse light with very different power spectrum properties and, more importantly,  there are negligible correlations between the maps used in the KAMM analysis and those of the ACS sources.

\section{IRAC data and results}

The GOODS IRAC data used for this investigation are described in KAMM2. Briefly: The GOODS IRAC observations of  the HDF-N and CDF-S fields cover regions roughly $16'\times10'$ in size. The coverage 
is obtained through two $10'\times10' $ mosaics (which overlap by $\sim4'$) that are obtained
at epochs $\sim6$ months apart (E1 and E2).
For this project the individual Basic Calibrated Data (BCD) frames were self--calibrated
using the method of Fixsen et al (2000), and mosaiced into images with $0\farcs6$ pixels. 
The self-calibration solves for fixed pattern detector offsets that may be incompletely removed
by the BCD pipeline (e.g. diffuse stray light), and for frame-to-frame offset variations.
The self-calibration does not remove the intrinsic large-scale power other than that due to (arbitrary) linear gradients, and is particularly useful for studies of the diffuse light fluctuations. 
Self-calibration cannot find a consistent solution if both epochs are combined,
because variation in the intensity, and especially the direction of the gradient, of the 
zodiacal light invalidate the assumption that the sky itself is a stable calibration source on 6-month timescales. 
Thus, our analyzed regions are four independent sub-fields of $583''\times583''$ in size with common 
coverage at both 3.6 and 4.5 \um. We refer to them below as HDFN-E1, HDFN-E2, CDFS-E1 and CDFS-E2. 

The assembled data were cleaned of resolved sources in two steps: 1) an iterative procedure was applied whereby each
iteration calculates the standard deviation ($\sigma$) of the
image, and then masks all pixels exceeding $N_{\rm cut}\ \sigma$
along with $N_{\rm mask}\times N_{\rm mask}$ surrounding pixels.
The procedure is repeated until no pixels exceed  $N_{\rm cut}\
\sigma$. The clipping parameters were fixed so that enough pixels
($\gsim70\%$) remain for robust Fourier analysis. As in the early DIRBE work (Kashlinsky \& Odenwald 2000) the blanked pixels were set to zero, and the resulting power spectrum was divided by the 
fraction of non-zero pixels in the image, 
thereby preserving the total power of the noise.
 2) At the second stage
we removed a model of the individual sources described in
KAMM1. This is done using a variant of the CLEAN
algorithm \cite{clean}. The process identifies the brightest pixel in the image, and then subtracts 
a scaled PSF at that location to remove a fixed fraction of the flux. This is repeated many thousands
of times until the components being subtracted are comparable to the noise level of the images.
Intermediate results are saved after each ``iteration'', consisting of the subtraction of 
$\sim 10^4$ components.
As progressively fainter sources are removed with each iteration, the shot noise from the remaining galaxies decreases as does their contribution to large scale power. 

Comparison of the different regions is most direct if the source cleaning 
is halted at the same specified level of $P_{\rm SN}$.
Fig.1 left shows the decrease of  $P_{\rm SN}$ with each iteration of the cleaning procedure for the four regions used in the study. The right panels in Fig.1 show the CIB fluctuations derived for the marked level of $P_{\rm SN}$ used in KAMM2. One can see the excess over the shot noise at scales $\gsim 0.5^\prime$. It is important to emphasize that the fluctuations have the large scale power spectrum, $P(q) \propto q^{-n}$, such that the amplitude of fluctuations is approximately flat to slowly rising with scale, corresponding to the effective index $n\gsim$2.

\section{ACS galaxies and their fluctuations}

We now evaluate the diffuse light  fluctuations that ACS galaxies would generate in the IRAC maps. For this we constructed synthetic maps over the same regions of the sky at 3.6 and 4.5 \um\ using the {\it HST}/ACS GOODS catalogs (v1.1z; Giavalisco et al. 2004). These 
were used to generate images of the observed $B$, $V$, $i$, and $z$-band 
galaxy distributions that are free of (a) small-scale noise 
and (b) intrinsically large-scale features,
whether due to a diffuse or unresolved astronomical component or
low-level instrumental calibration artifacts. The process for 
each ACS band was as follows: First, four blank images with the 
same scale and orientation of the IRAC images were generated 
for HDFN-E1,2 and CDFS-E1,2 regions. 
Then, for each cataloged source, the SExtractor (Bertin \& Arnouts 1996) parameters 
THETA\_IMAGE, A\_IMAGE, and B\_IMAGE were used to define a normalized 
Gaussian shape and orientation for each source, except for ``stellar'' sources  
(CLASS\_STAR $>$ 0.9) which were represented as a single pixel.
The intensity of each source was scaled to fit the MAG\_BEST magnitude, 
and each source was added to the initially blank image at the 
location designated by the ALPHA\_J2000 and DELTA\_J2000 parameters.
Finally, for comparison with the IRAC data, each of these images 
of the catalog sources was convolved with the 3.6 $\micron$ or 
4.5 $\micron$ PSF, as appropriate. For each of the four fields and at both 3.6 and 
4.5 \um, we prepared one image containing all ACS 
sources, and four others that included only ACS sources fainter than set AB magnitude limits, $[m_0,m_0+2,m_0+4,m_0+6]$ with $m_0=[21.8,21.8, 21.1, 20.6]$ in the ACS $B, V, i,z$-bands.

After the synthetic maps of ACS sources have been prepared, they were masked using the template derived from clipping (Step 1 above) of the IRAC data. Fig. 2 shows the fluctuations due to galaxies at all ACS wavelengths in the HDFN-E2 region (others are similar to within statistical uncertainties). 
Fluctuations of these {\it masked} maps show little difference out to a certain magnitude, which means that the clipping of IRAC data takes out galaxies very accurately for $m_{\rm AB}> $23--24 in $B, V, i, z$-band ACS filters. The power spectrum of the remaining sources is very different from that of the KAMM CIB fluctuations: it has effective index $n\lsim 0.5$ and is very similar to the spectrum of CIB fluctuations detected in the deep 2MASS data by Kashlinsky et al (2002), Odenwald et al (2003). The latter arises from galaxies fainter than $K_{\rm Vega} \simeq 19$ ($m_{\rm AB}\sim 21$ at 2.2 \um) which are located at $z\sim 1$ (Cirasuolo et al 2007) as are the bulk of the ACS galaxies. On the other hand the slope of the fluctuations in Fig. 1 (right) is described well by the concordance $\Lambda$CDM model at high $z$ (KAMM3). 
The differently sloped power spectra at large scales indicate that the resolved ACS galaxies are
unlikely contributors to the large--scale 3.6 and 4.5 \um\ fluctuations. The amplitude of the fluctuations at large scales is also small. Indeed the most that the ACS galaxies can contribute to the residual KAMM maps is constrained by the amount of the small-scale shot noise power measured to give fluctuations at a few arcsec of order $\sim 0.2$ \nwm2sr at both 3.6 and 4.5 \um. Scaling the numbers in Fig. 2 by the corresponding amount would then give contributions from the remaining ACS sources at arcminute scale well below the levels measured by KAMM and shown in Fig. 1 (right). This by itself implies that the KAMM CIB fluctuations cannot arise in the ACS galaxies. 

\section{Cross-correlating ACS and IRAC data}

If the ACS galaxies were to explain the KAMM signal there should be a strong correlation between the ACS source maps and KAMM maps, and their cross-correlation function must exhibit the same behavior at scales larger than the IRAC beam.
We computed both the correlation coefficients, $R_0\!=\!\langle \delta F_{\rm ACS}\delta F_{\rm KAMM}\rangle/\sigma_{\rm ACS}\sigma_{\rm KAMM}$, and the full cross-correlation matrix, $C(\theta)\!=\!\langle \delta F_{\rm ACS}(\vec{x})\delta F_{\rm KAMM}(\vec{x}+\vec{\theta})\rangle$ between the ACS and KAMM maps.

Fig 3 (left) shows a very good correlation between the IRAC sources removed by our modeling and the sources in the ACS catalog. On the other hand, only very small correlations (of order a few \%) remain between the residual KAMM maps, containing the fluctuations in Fig. 1 (right), and the ACS sources. These correlation coefficients also remain quite small as one sub-selects maps made only with progressively fainter ACS sources. 

To further test contributions of ACS galaxies to fluctuations on scales greater than the IRAC beam, we computed the correlation function $C(\theta)$ vs the separation angle $\theta$. In this representation, the contributions of any white
noise (such as shot noise and/or instrument noise) component to
$C(\theta)$ drop off very rapidly outside the beam and for the IRAC 3.6 \um\ channel contribute negligibly to the correlation function at $\theta$ greater than a few arcsec. This is best seen from the analog of the correlation coefficient at non-zero lag, defined as $R(\theta)\!\equiv\!C(\theta)/\sigma_{\rm ACS}\sigma_{\rm KAMM}$. This quantity is shown in the right panels of Fig. 3 using the example of the CDFS-E2 region. The mean square fluctuation on scale $\theta$ is given by the integral $\langle \delta F(\theta)^2\rangle$=$\frac{2}{\theta^2} \int_0^\theta C(\theta^\prime)\theta^\prime d\theta^\prime$ (e.g. Landau \& Lifshitz 1958). Hence, we evaluated from $R(\theta)$ a related quantity ${\cal R} (\theta)$=$\frac{2}{\theta^2} \int_0^\theta R(\theta^\prime)\theta^\prime d\theta^\prime$, shown in the lower right panels in Fig. 3. The correlations are negligible outside the IRAC beam, which means that, at most, the remaining ACS sources contribute to the shot-noise levels in the residual KAMM maps (as discussed by KAMM3), but not to the large scale correlations in Fig. 1 (right). 

\section{Conclusions}

Our analysis shows that the source-subtracted CIB fluctuations detected recently by KAMM in deep Spitzer images cannot originate in the optical galaxies seen in the GOODS ACS data. While the $B, V, i, z$-band galaxies are well correlated with those seen {\it and removed} in the 3.6 and 4.5 \um\ data, they correlate poorly with the the residual 3.6 and 4.5 \um\ background emission. These galaxies also exhibit a very different spatial power spectrum than the KAMM maps and the amplitude of their fluctuations is generally also low. 
Thus, as discussed in KAMM3, the distant ``ordinary" galaxies contribute to the shot noise component of the fluctuations, but their contribution to the clustering component is at most small. The amplitude of the KAMM source-subtracted CIB fluctuations requires significant CIB fluxes from objects fainter than the KAMM subtraction limit. Deep galaxy counts do not show signs of turning over at faint magnitudes, but their cumulative CIB levels saturate at magnitudes well below the KAMM removal threshold (Madau \& Pozzetti 2000, Fazio et al 2004). 

Whatever sources are responsible for the KAMM fluctuations, they are not present in the ACS catalog. 
There are two ways to reproduce this: 1) Since the ACS galaxies do not contribute to the  source-subtracted CIB fluctuations, the
latter must arise at $z\gsim$7.5 as is required by the Lyman break at rest $\sim$0.1 \um\ getting redshifted past the ACS $z$-band of peak wavelength $\simeq$0.85 \um. This would place the sources producing the KAMM signal within the first 0.7 Gyr of the evolution of the Universe and make conclusions of KAMM3 regarding their very low $M/L$ stronger. 2) Alternatively, the KAMM fluctuations would have to originate in lower $z$ galaxies which escaped the ACS GOODS source catalog either because they have low surface brightness or are below the catalog flux threshold, but at the same time generate diffuse light fluctuations of  higher amplitude and different slope than the populations already included there. In the latter case we can conservatively  estimate their expected luminosities, using the $z$-band $m_{\rm AB}\gsim$24-26.5 corresponding to the source size--dependent completeness limit of the ACS catalog (Giavalisco et al 2004). Such galaxies then would have to be very low-luminosity systems since e.g. for $m_{\rm AB}$=24 the in-band luminosity corresponding to $(B,z)$-band is $(4,2)\times 10^7 h^{-2}L_\odot$ at $z$=1 emitted at rest $0.22, 0.45$ \um.

We acknowledge support by NSF AST-0406587 and NASA Spitzer NM0710076 grants. 

%\texttt{\{thebibliography\}}%

\clearpage

\begin{figure}
\plotone{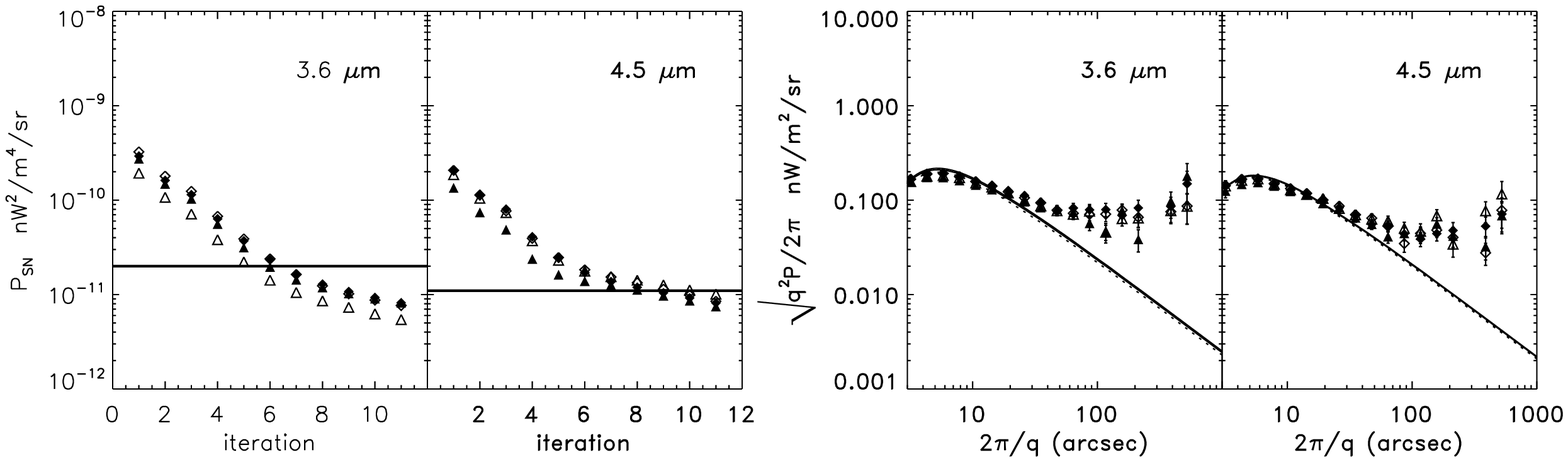} \caption{KAMM2 results for the $972\times 972$
0.6$^{\prime\prime}$ pixel fields. Triangles correspond to the
HDFN region, diamonds to CDFS region; open symbols to E1, filled
to E2. {\it Left}: Show decrease of the shot noise contribution to the power spectrum vs the iteration number of the KAMM source model. Horizontal lines show the levels reached in KAMM2, which are a factor of $\sim 2$ below those in KAMM1. {\it Right}: CIB fluctuations from the residual maps used in the present analysis at the shot-noise levels of KAMM2.} \label{fig:kamm}
\end{figure}

\clearpage

\begin{figure}
\plotone{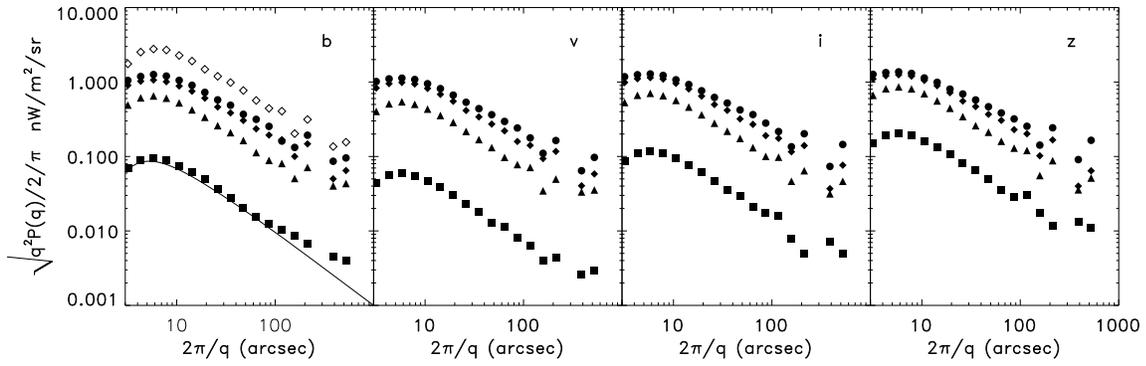} \caption{CIB fluctuations due to ACS galaxies
for the $972\times 972$ 0.6$^{\prime\prime}$ pixel field at
HDFN-E2 region. Filled circles correspond to ACS galaxies fainter
than $m_0$  with the mask defined by the clipping. Filled
diamonds, triangles and squares correspond to fluctuations produced
by sources fainter than $m_0+2,m_0+ 4, m_0+6$.
Open diamonds in the $B$-band panel show the fluctuations
produced by galaxies fainter than $m_{\rm AB}$=23.8 when the
clipping mask is not applied; the symbols show that such galaxies
were effectively removed from the ACS maps by clipping alone done
in KAMM. Solid line shows the shot noise slope corresponding to the faintest galaxies. It is clear that ACS galaxies are clustered, but the effective index of the clustering becomes closer to $n\sim$0 for fainter samples in general agreement with Fig.2 of Kashlinsky et al (2002).
} \label{fig:acs}
\end{figure}

\clearpage

\begin{figure}
\plotone{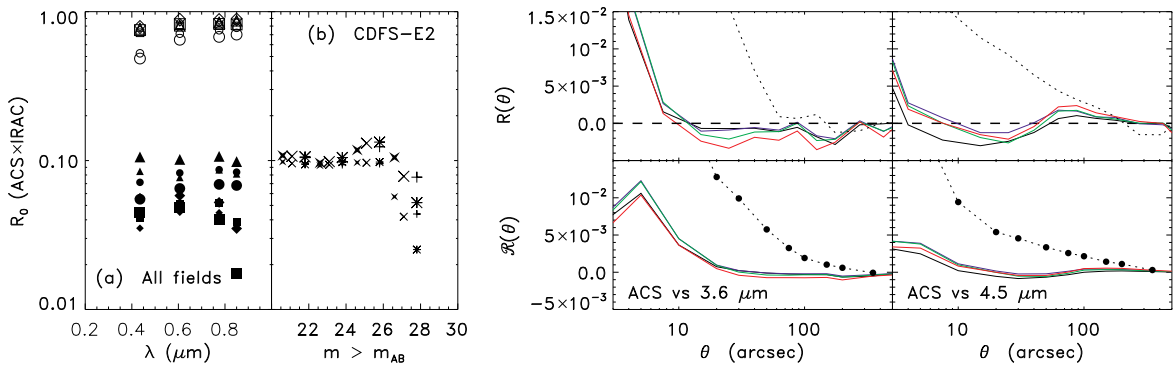} \caption{{\it Left}: Correlation coefficient between clipped/masked ACS and KAMM data. Large and small symbols correspond to the IRAC Ch 1 and Ch 2. (a) Shows the correlation coefficient when {\it all} ACS catalog sources are included. Open symbols correspond to correlations with the maps of the removed sources and filled symbols with the residual KAMM maps which contain the fluctuations shown in Fig. 1. Circles, diamonds, triangles, squares correspond to HDFN-E1, HDFN-E2, CDFS-E2, CDFS-E2. In some areas of the sky, such as CDFS-E2, a few bright sources (both galaxies and Galactic stars at $m_{\rm AB} \lsim 21$)  can noticeably reduce the correlation coefficient. (b) Shows the correlation coefficient with the ACS maps which include only sources {\it fainter} than the magnitude shown in the horizontal axis. Asterisks, pluses, X's and the four-pointed stars correspond to the maps using ACS $B,V,i,z$ sources. {\it Right}:  Black, blue, green, red solid lines show the dimensionless correlation function between the diffuse light in the ACS and KAMM maps for $B,V,i,z$-bands. Dotted line shows the dimensionless correlation function of the KAMM maps, $C_{\rm KAMM}(\theta)/\sigma_{\rm KAMM}^2$, which remains positive out to $\sim 100^{\prime\prime}$ and is better viewed when presented in log-log plots as in Fig. SI-4 of KAMM1. The dashed horizontal line marks $R=0$.  The lower panels show a more direct measure of the fluctuations on a given scale, ${\cal R}(\theta)=\frac{2}{\theta^2} \int_0^\theta R(\theta^\prime)\theta^\prime d\theta^\prime$.
} \label{fig:acs_vs_kamm}
\end{figure}

\clearpage

\end{document}